\documentclass[prl,showpacs,showkeys,nofootinbib,twocolumn,floatfix]{revtex4}

\usepackage{graphicx}  %
\usepackage{amsmath}
\usepackage{bm}  %
\usepackage{ulem}

\newcommand{\bea}{\begin{eqnarray}}
\newcommand{\eea}{\end{eqnarray}}
\newcommand{\beq}{\begin{equation}}
\newcommand{\eeq}{\end{equation}}
\newcommand{\bqa}{\begin{eqnarray}}
\newcommand{\eqa}{\end{eqnarray}}

\def\mqo2{{\!\!\!}}

\begin{document}

\title{Dense Axion Stars}

\author{Eric Braaten}
\email{braaten@mps.ohio-state.edu}
\affiliation{Department of Physics,
         The Ohio State University, Columbus, OH\ 43210, USA\\}

\author{Abhishek Mohapatra}
\email{mohapatra.16@buckeyemail.osu.edu}
\affiliation{Department of Physics,
         The Ohio State University, Columbus, OH\ 43210, USA\\}

\author{Hong Zhang}
\email{zhang.5676@osu.edu}
\affiliation{Department of Physics,
         The Ohio State University, Columbus, OH\ 43210, USA\\}

\date{\today}

\begin{abstract}
If the dark matter particles are axions,
gravity can cause them to coalesce into axion stars, 
which are stable gravitationally bound  systems of axions.
In the previously known solutions for axion stars,
gravity and the attractive force between pairs of axions are 
balanced by the kinetic pressure. 
The mass of these dilute axion stars cannot exceed a critical mass,
which is about $10^{-14} M_\odot$ if the axion mass is $10^{-4}$~eV.
We study axion stars using  a simple
approximation to the effective potential of the 
nonrelativistic effective field theory for axions.
We find a  new branch of dense axion stars
in which gravity is balanced by the mean-field pressure of the axion Bose-Einstein condensate.
The mass on this branch ranges from 
about $10^{-20} M_\odot$ to about $M_\odot$.  
If a dilute axion star with the critical mass accretes additional axions and collapses,
 it could produce a bosenova, leaving a dense axion star as the remnant.

\end{abstract}

\smallskip
\maketitle

{\bf Introduction}.
The particle nature of the  dark matter of the universe
remains one of the greatest mysteries in contemporary physics.
One of the most strongly motivated possibilities from a particle physics perspective
 is the {\it axion} \cite{Kim:2008hd},
which is the pseudo-Goldstone boson
associated with a $U(1)$ symmetry that solves the strong $CP$ problem of QCD.
The axion is a spin-0 particle with a very small mass
and extremely weak self-interactions.
Axions can be produced in the early universe by a vacuum misalignment mechanism \cite{Kim:2008hd}.
The resulting axions are extremely nonrelativistic and have huge occupations numbers,
so they are naturally described by a classical field.
Sikivie and collaborators have pointed out that gravitational interactions  
can continually rethermalize axion dark matter \cite{Sikivie:2009qn,Erken:2011vv}.
They can therefore form a Bose-Einstein condensate (BEC)  that remains locally 
in the lowest-energy accessible  state.

A stable gravitationally bound configuration of axions is called an {\it axion star}.
In the ground state of an axion star, the axions are in a BEC.
In the known solutions for axion stars,
the attractive forces from gravity and from the 4-axion interaction
are balanced by the kinetic pressure.
We refer to them as {\it dilute axion stars},
because the number density remains small enough that 
6-axion and higher interactions are negligible.
There is a critical mass for the dilute axion star beyond which the kinetic pressure 
is unable to balance the attractive forces. 
The fate of an axion star that exceeds the critical mass 
and begins to collapse has not been established.
The possibilities for the remnant after the collapse include
\begin{itemize}
\item
{\it black hole}. 
Axion self-interactions may be too weak to prevent the collapse 
of the axion star to a configuration smaller than its Schwarzschild radius. 
\item
{\it dilute axion star} with smaller mass.
The decrease in mass could come from radiation of axion waves or from a {\it bosenova}, in which
inelastic axion reactions amplified by the increasing number density during the collapse
produce a burst of relativistic axions.
(An analogous phenomenon has been observed in the collapse of a BEC of 
ultracold atoms \cite{JILA-bosenova}.)
\item
{\it dense axion star},  
in which the number density becomes so high that
multi-axion interactions affect the balance of forces (provided such solutions exist).
\end{itemize}

In this paper, we study axion stars using  a simple approximation to the effective potential of the 
nonrelativistic effective field theory for axions.
We find a new branch of dense axion stars
in which the attractive force of gravity is balanced by the mean-field pressure
of the axion BEC. The mass of a dense axion star can be 
many orders of magnitude smaller or larger than the critical mass of a dilute axion star.

{\bf Relativistic axion field theory.}
Axions can be described by a relativistic quantum field theory with a real scalar field $\phi(x)$.
The Hamiltonian has the form
\begin{equation}
{\cal H} = \tfrac{1}{2}\dot{\phi}^2
+\tfrac{1}{2} \nabla \phi \cdot \nabla \phi
+ {\cal V}(\phi) .
\label{H-phi}
\end{equation}
The potential ${\cal V}$ is produced by nonperturbative QCD effects, and depends 
on quark mass ratios  \cite{diCortona:2015ldu}.
It is commonly approximated by
\begin{equation}
{\cal V}(\phi)  = 
m^2 f^2\left[ 1 - \cos(\phi/f) \right],
\label{V-phi}
\end{equation}
where $m$ is the mass of the axion and $f$ is the axion decay constant.
In this paper, we assume that 
${\cal V}$ in Eq.~\eqref{V-phi} is an adequate approximation for the relativistic potential.

The product $m f$ of the mass  of the axion  and its decay constant  is  
$(8 \times 10^7~{\rm eV})^2$ \cite{Kim:2008hd}.
Astrophysical and cosmological constraints restrict $f$
to the window between about $5 \times 10^{17}$~eV
and about $8 \times 10^{21}$~eV  \cite{Kim:2008hd}.
The window for $m$ is therefore
from about $10^{-6}$~eV to about $10^{-2}$~eV.

{\bf Nonrelativistic axion effective field theory.}
Axions whose kinetic energies are much smaller than $m$ can be described by a 
nonrelativistic effective field theory with a complex scalar field $\psi(\bm{r},t)$.
We refer to this effective theory as {\it axion EFT}. 
The number of axions, $N = \int \!d^3 r\, \psi^*\psi$, is conserved in axion EFT.
The effective Hamiltonian has the form
\begin{equation}
{\cal H}_{\rm eff} = 
\frac{1}{2m} \nabla \psi^* \cdot \nabla \psi
+ {\cal V}_{\rm eff}(\psi^* \psi) .
\label{Heff-psi}
\end{equation}
The effective potential ${\cal V}_{\rm eff}$ is the generator of $n \to n$ axion scattering vertices.
In the case of an axion BEC, $ {\cal V}_{\rm eff}$ also gives 
the mean-field energy of the condensate as a function of its number density $\psi^* \psi$.
The coefficients in the expansion of ${\cal V}_{\rm eff}$ 
 in powers of $\psi^* \psi$ can be derived by matching low-energy 
scattering amplitudes at tree level in the relativistic theory and in axion EFT \cite{Braaten:2016kzc}.

A  simple approximation to ${\cal V}_{\rm eff}$ can be obtained by 
a nonrelativistic reduction of the relativistic Hamiltonian in Eq.~\eqref{H-phi}.
After inserting
\begin{equation}
\phi(\bm{r},t)  = \frac{1}{\sqrt{2m} }
\left[ \psi(\bm{r},t) e^{-imt} + \psi^*(\bm{r},t) e^{+imt} \right],
\label{phi-psi}
\end{equation}
dropping  terms with a rapidly oscillating phase factor $e^{inmt}$, 
where $n$ is a nonzero integer,
and also dropping terms with time derivatives of $\psi$,
we obtain the effective Hamiltonian in Eq.~\eqref{Heff-psi} with the effective potential
\begin{equation}
{\cal V}_{\rm eff}( \psi^* \psi) = \tfrac12 m \psi^* \psi
+ m^2  f^2 [ 1 - J_0(\hat n^{1/2}) ],
\label{Veff-psi}
\end{equation}
where  $\hat n = 2 \psi^* \psi/mf^2$ and $J_0(z)$ is a Bessel function.
The first term arises from the $\dot{\phi}^2$ term in Eq.~\eqref{H-phi}.
The second term was derived previously in Ref.~\cite{Eby:2014fya}.
At small $\hat n$, the second term approaches $\frac12 m \psi^* \psi$.
At large $\hat n$, it approaches $m^2 f^2$, oscillating around that value with a decreasing amplitude.
The expansion of ${\cal V}_{\rm eff}$ in Eq.~\eqref{Veff-psi} in powers of $\psi^* \psi$ 
agrees with the exact effective potential for axion EFT through order $(\psi^* \psi)^2$ \cite{Braaten:2016kzc}.
In this paper, we  assume that ${\cal V}_{\rm eff}$ in Eq.~\eqref{Veff-psi} 
is an adequate approximation for the effective potential.

{\bf Axion stars.}
A {\it boson star} consists of a large number of bosons  that are gravitationally bound.
Boson stars consisting of  bosons with no self-interactions and
with gravitational interactions described by general relativity were first studied
numerically by Kauf \cite{Kaup:1968} and by Ruffini and Bonazzola \cite{RB:1969}. 
The critical mass $M_*$ beyond which the boson star is unstable to gravitational collapse
was first determined accurately by Breit, Gupta and Zaks \cite{Breit:1983nr}:
$M_* = 0.633/(Gm)$,
where $G$ is Newton's gravitational constant and $m$ is the mass of the boson.
If $m= 10^{-4\pm 2}$~eV, 
the critical mass is $8.4 \times 10^{-7\mp 2}~M_\odot$, 
where $M_\odot$ is the mass of the sun.
(Here and below, the upper and lower error bars in an exponent correspond to increasing 
and decreasing $m$ by two orders of magnitude from $10^{-4}$~eV.)

An {\it axion star} is a boson star consisting of axions,
a possibility first considered by Tkachev \cite{Tkachev:1991ka}.
 Axion stars  with axion self-interactions given by 
expanding ${\cal V}$  in Eq.~\eqref{V-phi} to order $\phi^6$
and with gravitational interactions described by general relativity were studied numerically
by Barranco and Bernal \cite{Barranco:2010ib}.
They found solutions that correspond to dilute axion stars.
The axions in a dilute axion star are nonrelativistic and
their gravitational interactions are accurately described by Newtonian gravity.
The gravitational energy is
\begin{equation}
E_{\rm gravity}  = 
- \frac{G m^2}{2} \int \!\!d^3r_1 \!\! \int \!\!d^3r_2
\frac{ \psi^* \psi(\bm{r}_1) \psi^* \psi(\bm{r}_2)}{|\bm{r}_1 - \bm{r}_2|}.
\label{Egravity-psi}
\end{equation}
Axion stars  with axion self-interactions  given by 
expanding ${\cal V}$ in Eq.~\eqref{V-phi} to order $\phi^4$
and with gravitational interactions described by Newtonian gravity were studied
by Chavanis and Delfini \cite{Chavanis:2011zm}.
Their result for the critical mass $M_*$ of the dilute axion star
above which there are no stable spherically symmetric solutions is 
\begin{equation}
M_* = 10.15~ f/\sqrt{Gm^2}.
\label{Mmax-axion}
\end{equation}
If  $m= 10^{-4\pm 2}$~eV,
the critical mass is $6 \times 10^{-14\mp 4}\,M_\odot$,
which corresponds to   $7 \times 10^{56 \mp 6}$ axions.
The radius of the dilute axion star decreases as $M$ increases.
The radius $R_{99}$ that encloses 99\% of the axions 
is $9.9/GMm^2$ for small $M$,
and it decreases to
$0.55/(Gm^2f^2)^{1/2}$  at the critical mass  \cite{Chavanis:2011zm}.
This minimum radius is $3 \times 10^{-4}R_\odot$, 
where $R_\odot$ is the radius of the sun.

The equations that describe a static  BEC of axions
can be obtained by considering an axion field $\psi(\bm{r},t)$ whose time-dependence is 
$\psi(\bm{r}) \exp(-i \mu t)$, where $\mu$ is the chemical potential.
They can be expressed conveniently as coupled equations for $\psi(\bm{r})$
and the gravitational potential $\phi(\bm{r})$:
\begin{subequations}
\begin{eqnarray}
\nabla^2 \psi &=& -2m \big[ \mu - ({\cal V}_{\rm eff}'(\psi^* \psi)- m) - m \phi \big] \psi,
\label{static-psi}
\\
\nabla^2 \phi &=& 4 \pi Gm \psi^* \psi.
\label{static-phi}
\end{eqnarray}
\label{static-psiphi}%
\end{subequations}
For spherically symmetric solutions, $\psi(r)$ and $\phi(r)$ are functions of $r$ only.
It is convenient to take the central number density $n_0 = |\psi(0)|^2$ 
as an input parameter.  If $|\psi(0)| = n_0^{1/2}$
is specified, the other boundary condition at the origin is
$\psi'(0) = 0$ and the boundary conditions as $r \to \infty$ are $\psi(r) \to 0$ and $\phi(r) \to 0$.
These boundary conditions determine the eigenvalue $\mu$.

\begin{figure}[t]
\centerline{ \includegraphics*[width=8cm,clip=true]{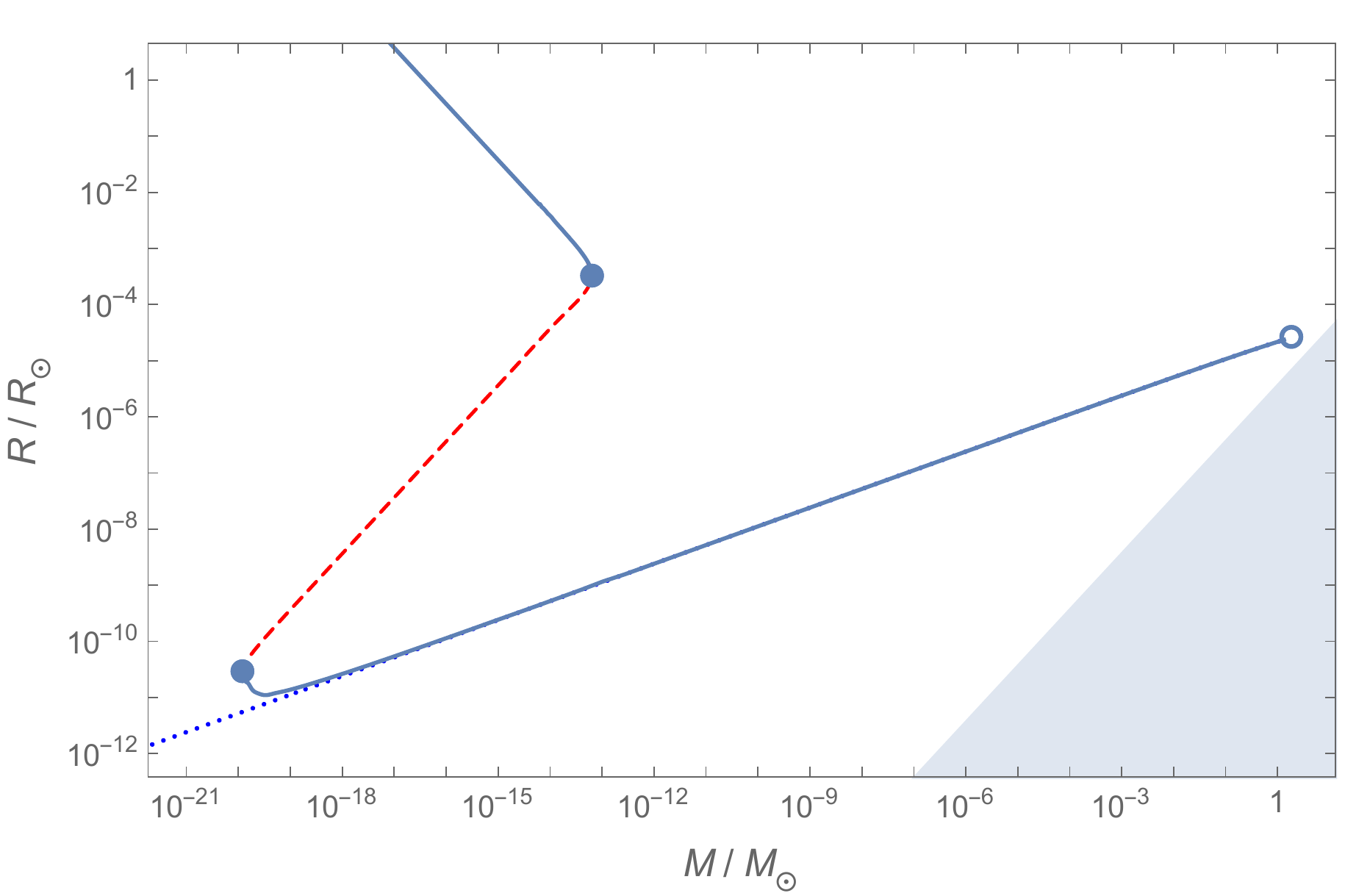} }
\vspace*{0.0cm}
\caption{Radius $R_{99}$ of an axion star (normalized to $R_\odot$)
as functions of its mass $M$ (normalized to $M_\odot$) for $m = 10^{-4}$~eV.
The curves for the radius are  stable dilute axion stars (upper solid line),
unstable axion stars (dashed line), stable dense axion stars (lower solid line),
and the Thomas-Fermi approximation (dotted line).
The shaded region is below the Schwarzchild radius $2GM/c^2$.
}
\label{fig:RvsM}
\end{figure}

We first consider dilute axion stars, in which
the axion number density $\psi^* \psi$  is always small compared to $m f^2$.
The solution of Eqs.~\eqref{static-psiphi} gives the same results whether we use
the effective potential in Eq.~\eqref{Veff-psi}  or we truncate it
after the $(\psi^* \psi)^2$ term so that
${\cal V}_{\rm eff}' - m = -\psi^* \psi/(8f^2)$.
The mass-radius relation for dilute axion stars is shown in 
the upper left corner of Fig.~\ref{fig:RvsM}.  
As the dimensionless central density $\hat n_0 = 2|\psi(0)|^2/mf^2$ increases,
 the radius $R_{99}$ decreases
and the mass $M$ increases until it reaches the critical mass $M_*$ in Eq.~\eqref{Mmax-axion}
at $\hat n_0 = 1.3 \times 10^{-14 \mp 4}$.
Beyond this critical point, the solutions are unstable 
under perturbations that increase the central density \cite{Chavanis:2011zm}.

We now consider dense axion stars, in which $\psi^* \psi$ can be comparable 
to or larger than $m f^2$.
In this case, one must use an approximation to ${\cal V}_{\rm eff}$
that includes all orders in $\psi^*\psi$,
such as the effective potential in Eq.~\eqref{Veff-psi}.
As illustrated in Fig.~\ref{fig:RvsM},
the unstable solutions for a dilute axion star can be continued 
from the critical point at $M_*$ to smaller values of $M$
where ${\cal V}_{\rm eff}$ can no longer be truncated after the $(\psi^* \psi)^2$ term.
The mass decreases to a minimum value $M_*'$ at a second critical point
where $\hat n_0 = 13$, and then $M$ begins increasing.
At both critical points, the slope of the chemical potential $\mu$
as a function of $M$ is infinite.  By Poincare's theory of linear series of equilibria,
the number of unstable modes must change by one at each critical point \cite{Katz:1978}.
We have not excluded the possibility that the number of unstable modes changes to two
after the second critical point, but we assume it changes to zero.
This assumption is supported by the Thomas-Fermi approximation described below.
The solutions beyond the second critical point are then stable dense axion stars,
in which the mean-field pressure of the axion BEC balances 
the force of gravity and the kinetic pressure.
For $m= 10^{-4\pm2}$~eV, the second critical mass $M_*'$ is $1.2 \times 10^{-20 \mp6}M_\odot$,
which corresponds to $1.3 \times 10^{50 \mp 8}$ axions. 
The radius $R_{99}$ at the critical point is $2.6 \times 10^{-11 \mp 2}R_\odot$,
which is equal to $9.2(\hbar /mc)$. 

{\bf Thomas-Fermi approximation.}
Beyond the second critical point,
the accurate numerical solution of Eqs.~\eqref{static-psiphi}
becomes increasingly challenging as $n_0$ increases further.
The mean-field energy also becomes increasingly large compared to the kinetic energy.
The solution can therefore be simplified using 
the {\it Thomas-Fermi approximation}, in which the kinetic energy is neglected.
This approximation provides an excellent description of BECs 
of ultracold atoms when the number of atoms is large enough that 
their interaction energy is comparable to the trapping energy \cite{BP:1996}.
It has been applied previously by Wang to boson stars in which the 
interaction between a pair of bosons is  repulsive \cite{Wang:2001wq}.
In the Thomas-Fermi approximation,
the $\nabla^2 \psi$ term on the left side of Eq.~\eqref{static-psi} is neglected.
For the effective potential $ {\cal V}_{\rm eff}$ 
in Eq.~\eqref{Veff-psi}, Eq.~\eqref{static-psi} reduces to
\begin{equation}
0 = \left[ \mu - m \big(J_1(\hat n^{1/2})/\hat n^{1/2} -  \tfrac12 \big) - m \phi \right] \psi,
\label{TF-psi}
\end{equation}
where $J_1(z)$ is a Bessel function.
This equation either requires $\psi = 0$ or it determines  $\phi$ in terms of $\psi$.
Inserting the expression for $\phi$ into Eq.~\eqref{static-phi}
gives a self-contained second-order differential equation for $\psi$.
Given the central density $n_0$ and the boundary condition $\psi'(0)=0$,
that differential equation  can be integrated to obtain $\psi(r)$ as a function of $r$.
If $\psi(r)$ does not cross zero, there is no solution 
to  Eqs.~\eqref{static-psiphi} for that central density.
If $\psi(r)$ crosses 0 at some radius $R_{\rm TF}$, there is a solution.
The Thomas-Fermi approximation  is $\psi(r)$ for $r<R_{\rm TF}$ and 0 for $r>R_{\rm TF}$.
We refer to $R_{\rm TF}$ as the {\it Thomas-Fermi radius}.
This approximation is continuous, but its first derivative has a discontinuity at $r=R_{\rm TF}$.
In Fig.~\ref{fig:RvsM}, 
the Thomas-Fermi approximation has been used to extend the branch of dense axion stars
up to a mass of $1.9\, M_\odot$, 
beyond which there is no solution within this approximation.
 The relation between $R_{\rm TF}$ and $M$ is independent of $m$ for fixed $mf$.
It is accurately approximated by a power law:
$R_{\rm TF}/R_\odot = 2.1 \times 10^{-5} (M/M_\odot )^{0.305}$.
The Thomas-Fermi approximation has solutions for arbitrarily small $M$. 
Its extrapolation to $M$ below its range of validity is  shown in Fig.~\ref{fig:RvsM}.
The region where $R$ is smaller than the Schwarzchild radius $2 G M/c^2$ 
of the axion star is shown in Fig.~\ref{fig:RvsM} as a shaded region.
Since $R_{\rm TF}$ is much larger than $2 G M/c^2$
over most of the range of a dense axion star,
Newtonian gravity is a good approximation.

\begin{figure}[t]
\centerline{ \includegraphics*[width=8cm,clip=true]{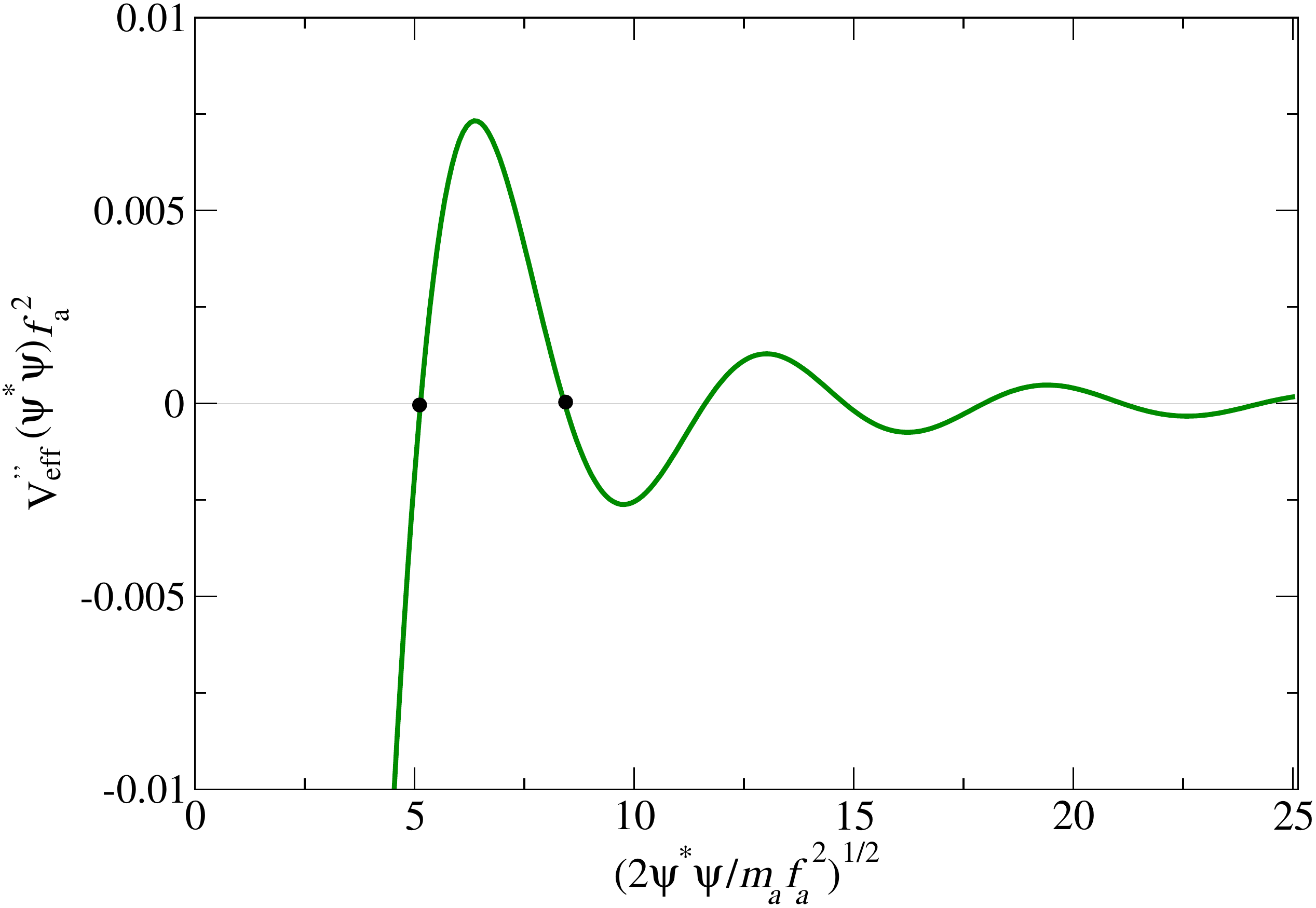} }
\vspace*{0.0cm}
\caption{Second derivative of the effective potential 
with respect to its argument $\psi^*\psi$  as a function of $|\psi|$.
The left and right dots mark values of $|\psi|$ near those at the center of the dense axion star  
with the smallest mass and the largest mass, respectively.
}
\label{fig:Veff''}
\end{figure}

{\bf Other branches.}
The existence of dense axion star solutions can be understood 
by considering the balance of forces in the Thomas-Fermi approximation.
The forces are the negatives of the gradients of the energies 
inside the square brackets in Eq.~\eqref{static-psi}.  The condition for the balance of forces is
\begin{equation}
{\cal V}_{\rm eff}''(\psi^* \psi) \frac{d\ }{dr}(\psi^* \psi) + m \frac{d\ }{dr}\phi = 0.
\label{F-balance}
\end{equation}
If $\psi^* \psi$ is a positive decreasing function of $r$ 
and $\phi$ is a negative increasing function of $r$,
the balance of forces requires ${\cal V}_{\rm eff}''$ to be positive.
In a region of $r$ where ${\cal V}_{\rm eff}''>0$,
the repulsive mean-field pressure of the axion BEC 
can balance the attractive gravitational force.
In a region of $r$ where ${\cal V}_{\rm eff}''<0$,
the mean-field pressure is attractive and 
the two attractive forces must be balanced by the kinetic pressure of the axions.
In Fig.~\ref{fig:Veff''}, ${\cal V}_{\rm eff}''$ is shown as a function of the 
dimensionless number density variable $\hat n^{1/2}$.
The second derivative is positive between the two dots
where $\hat n^{1/2}$ is 5.14 and 8.42. 
For most of the range of stable dense axion stars in Fig.~\ref{fig:RvsM},
$\hat n_0^{1/2}$ is in this range.
As $r$  increases, $\hat n^{1/2}$ decreases  steadily,
with the repulsive mean-field pressure balancing gravity, 
until it  approaches 5.14.
As $r$ continues to increase, a large kinetic pressure is required to balance 
gravity and the attractive mean-field pressure.  The large gradient of $\psi$
results in a rapid decrease of $\psi$ to zero.

In Fig.~\ref{fig:Veff''}, there are other intervals of $\hat n^{1/2}$ in which ${\cal V}_{\rm eff}''$ is positive.
For each of these intervals, there may be another branch of stable dense axion stars
in which $\hat n_0^{1/2}$  is in this interval.
As $r$  increases from 0 to $\infty$, $\hat n^{1/2}$ must decrease to 0.
It can decrease steadily with $r$ through the values of $\hat n^{1/2}$
for which ${\cal V}_{\rm eff}''>0$,
but $\psi(r)$ must decrease precipitously with $r$ through the values of $\hat n^{1/2}$
for which ${\cal V}_{\rm eff}''<0$. 
The Thomas-Fermi approximation is not applicable,
because the kinetic pressure is important in regions where ${\cal V}_{\rm eff}''<0$.

{\bf Discussion.}
In our solutions for  dense axion stars, we used the simple model  in  Eq.~\eqref{V-phi} 
for the relativistic axion potential
${\cal V}$ and  we approximated the effective potential
${\cal V}_{\rm eff}$ of axion EFT by Eq.~\eqref{Veff-psi}.
To obtain more accurate solutions, it would be necessary to use 
a more accurate relativistic axion potential   \cite{diCortona:2015ldu}.
It would also be important to develop a systematically improvable 
approximation to the effective potential ${\cal V}_{\rm eff}$ 
of axion EFT \cite{Braaten:2016kzc}. 
A more accurate approximation for ${\cal V}_{\rm eff}$ would certainly change
the quantitative properties of dense axion stars,
and it could even affect their existence.

An important issue in axion dark matter is whether the BEC 
can be coherent over galactic and cosmological  distances, 
as argued by Sikivie and collaborators  \cite{Sikivie:2009qn,Erken:2011vv}.
Guth, Hertzberg, and Prescod-Weinstein have recently argued that 
the attractive interaction between a pair of axions prevents the development of
long-range correlations in the axion BEC \cite{Guth:2014hsa}.
This suggests that the largest possible coherence length is 
roughly the radius $10^{-4}R_\odot$
of a dilute axion star with the critical mass. 
Since dense axion stars  have smaller radii,
their existence does not modify the conclusions of  Ref.~\cite{Guth:2014hsa}.

Dense axion stars with sufficiently large masses
could be observed through gravitational microlensing.
The possibility that most of the dark matter in the Milky Way is made up of massive compact halo objects
(such as axion stars) with masses in the range between $2 \times 10^{-9}M_\odot$
and $15 M_\odot$ has been excluded \cite{EROS-2,Griest:2013esa}.
Thus most of axion dark matter must be in the form of gases of axions
or dilute axion stars or dense axion stars 
with mass less than $2 \times 10^{-9}M_\odot$.
To  obtain quantitative constraints on axions from microlensing,
it would be necessary to know the fraction of axions
that are bound in axion stars as well as the mass distributions of dilute axion stars 
and of dense axion stars.

An interesting issue is the ultimate fate of a dilute axion star with the critical mass $M_*$
that accretes additional axions and begins to collapse.
Before this work, the most likely possibility for the remnant after the collapse 
seemed to be a black hole.  The alternative possibility was
a dilute axion star with a smaller mass and a larger radius, 
but it is hard to imagine a plausible path to that large dilute axion star 
from the tiny collapsing axion star.
Solutions for stable dense axion stars suggest 
an alternative possibility for the remnant after the collapse:
a dense axion star with mass smaller but of order $M_*$. 
The decrease in radius by more than 5 orders of magnitude during  the collapse
could amplify inelastic axion reactions, such as the scattering of 4 axions 
from the BEC into 2 relativistic  axions,
producing a bosenova \cite{JILA-bosenova}.
The relaxation to a static dense axion star would also involve the radiation of axion waves.
It should be possible to determine the fate of a collapsing dilute axion star
definitively by solving the time-dependent classical field equations of axion EFT.

If the dark matter consists of axions, the formation of dense axion stars 
in the early universe may be inevitable.  
Spacial fluctuations in the vacuum misalignment of the axion field naturally produce
gravitationally bound �miniclusters� of axions with masses comparable to the
critical mass $M_*$ of a dilute axion star \cite{Hogan:1988mp,Kolb:1993zz}.
Gravitational thermalization will drive a minicluster towards an axion star
and ensure that the axion star remains a BEC as it accretes more  axions.  
If accretion of axions increases the mass of the dilute axion star to above $M_*$,
it may collapse into a dense axion star with a smaller mass of order $M_*$.
The resulting mass distribution of dense axion stars could be modified by the 
subsequent accretion of additional axions.
The existence of dense axion stars would
open up many new possibilities in the search for axion dark matter.


\begin{acknowledgments}
This research was supported in part by the
Department of Energy under the grant DE-SC0011726
and by the National Science Foundation under the grant PHY-1310862.
We thank J.~Beacom, S.~Raby, and  B.~Shi for useful discussions.
\end{acknowledgments}

\end{document}